# Electronic structure of the iron-based superconductor LaOFeP

D. H. Lu[1], M. Yi[1], S.-K. Mo[1,2], A. S. Erickson[3], J. Analytis[3], J.-H. Chu[3], D. J. Singh[4], Z. Hussain[2], T. H. Geballe[3], I. R. Fisher[3], and Z.-X. Shen[1]

[1]Department of Physics, Applied Physics, and Stanford Synchrotron Radiation Laboratory, Stanford University, Stanford, CA 94305, USA

[2]Advanced Light Source, Lawrence Berkeley National Lab, Berkeley, CA 94720, USA

[3]Geballe Laboratory for Advanced Materials and Department of Applied Physics, Stanford University, Stanford, California 94305-4045, USA

[4]Materials Science and Technology Division, Oak Ridge National Laboratory, Oak Ridge, Tennessee 37831-6114, USA

**The recent discovery of superconductivity in the so-called iron-oxypnictide family of compounds[1-9] has generated intense interest. The layered crystal structure with transition metal ions in planar square lattice form and the discovery of spin-density-wave order near 130K[10,11] seem to hint at a strong similarity with the copper oxide superconductors. A burning current issue is the nature of the ground state of the parent compounds. Two distinct classes of theories have been put forward depending on the underlying band structures: local moment antiferromagnetic ground state for strong coupling approach[12-17] and itinerant ground state for weak coupling approach[18-22]. The local moment magnetism approach stresses on-site correlations and proximity to a Mott insulating state and thus the resemblance to cuprates; while the**

**latter approach emphasizes the itinerant electron physics and the interplay between the competing ferromagnetic and antiferromagnetic fluctuations. Such a controversy is partly due to the lack of conclusive experimental information on the electronic structures. Here we report the first angle-resolved photoemission spectroscopy (ARPES) investigation of LaOFeP ($T_c$ = 5.9 K), the first reported iron-based superconductor[2]. Our results favor the itinerant ground state, albeit with band renormalization. In addition, our data reveal important differences between these and copper based superconductors.**

Fig. 1 compares the angle-integrated photoemission spectrum (AIPES) with the density of states (DOS) obtained from the LDA band structure calculations. It is important to note that the peak near the Fermi level ($E_F$) is as strong as the valence band (VB) peak, in sharp contrast with the typical VB spectrum of cuprates as shown in the inset of Fig. 1**a**. The VB spectrum of cuprates is characterized by a weak feature near $E_F$ on top of a broad VB peak, consistent with the doped Mott insulator picture. This clear disparity between the iron-based superconductor and cuprates suggests that itinerant rather than Mott physics is a more appropriate starting point for the iron-based superconductor, at least for LaOFeP discussed in this paper. Our data also contrast with some recent AIPES data[24,25] obtained from polycrystalline samples that only show a very small peak near $E_F$ on top of a large VB peak, which is reminiscent of VB spectrum of cuprates. This difference may be caused by the surface quality of polycrystalline samples as often the case for oxides[26]. In balance, our data do not support theoretical models assuming strongly antiferromagnetic ground states, at least those currently being formulated, albeit for the LaOFeAs system[12-15], as there is no evidence in our spectra for exchange splitting of the Fe $d$ states, and agreement between our VB spectrum and those calculated DOS is poorer.

More detailed information can be obtained from angle-resolved data. To understand the seemingly complex multi-bands electronic structure, we overlay the LDA band structures on top of our data (Fig. 2). A quantitative agreement can be found between the ARPES spectra and the calculated band dispersions after shifting the calculated bands up by 0.11 eV and then renormalizing by a factor of 2.2. Note that the values of $E_F$ shift and band renormalization factor are chosen to obtain the best match of the two higher binding energy bands at the Γ-point. While the renormalized bands using this set of parameters fit the bands near Γ very well, the match near the X-point and the M-point is less perfect. This suggests that different bands may have slightly different renormalization effects. Nevertheless, the overall degree of agreement between the experiments and the calculations is rather remarkable as nearly all features presented in our data have corresponding bands in the calculations, indicating that the LDA calculations assuming an itinerant ground state capture the essence of the electronic structures of this system. This again suggests that the iron-based superconductors, at least LaOFeP, are different from cuprate superconductors. We also note that the measured dispersions show no similarity with the band structure calculations of LaOFeAs that assume an antiferromagnetic ground state[12-15].

To extract more information from ARPES spectra, a simple momentum-distribution-curve analysis was applied to the high symmetry cuts. A Fermi velocity ($v_F$) of 1.0±0.2 eV·Å is obtained for all three bands individually. For comparison, the values extracted from the LDA calculations, after taking into account the $E_F$ shift, are 1.5/1.7, 1.4, and 2.4/3.5 eV·Å for $\varGamma_1$, $\varGamma_2$, and $M$ bands, respectively. Note that two different numbers are given for both $\varGamma_1$ and $M$ bands since each contains two nearly degenerate bands. This observation demonstrates that the renormalization effects are different for different bands, as anticipated above, indicating that correlation effects are appreciable and not isotropic. However, these $v_F$ renormalization values as well as the factor of 2.2 total bandwidth rescaling are comparable to those of $Sr_2RuO_4$, which is a correlated Fermi liquid and is reasonably well described by theories using itinerant band structure as the starting point[27]. The corresponding free electron band masses $m^*$ extracted from our data are 1.4±0.3, 4.6±0.5, and 1.3±0.3 $m_e$ for $\Gamma_1$, $\Gamma_2$, and $M$ bands, respectively. It is interesting to note that the magnetic susceptibility enhancement compared with the bare band structure DOS is nearly a factor of 6[18], indicating either a lower energy scale renormalization or a strong Stoner renormalization. In this regard, we do not observe any apparent low energy kink in the dispersion near 50 meV, which is a universal feature in cuprates[28].

Fig. 3 displays the energy-distribution-curves (EDC's) along the same high symmetry cuts as shown in Fig. 2. Taking a close look at these EDC's, we find that there is no evident pseudogap within our experimental uncertainty in all three bands crossing the Fermi level, contrasting to the ubiquitous pseudogap observed in underdoped cuprates. The absence of the pseudogap, therefore, marks an important difference between this new iron-based superconductor and cuprates. This finding contradicts the recent report of a 20 meV pseudogap in LaOFeP from AIPES[29]. The difference can be attributed to either a poor surface quality of the polycrystalline samples used for that measurement, where previous experience indicates potential problems associated with impurities[26], or distortion of the AIPES result by states away from $k_F$. ARPES from single crystalline samples is much better suited to address the pseudogap issue by directly measuring the states near $k_F$. The same and another AIPES experiments also indicated pseudogap effects with energy scales of 20 and 100 meV in polycrystalline $La[O_{1-x}F_x]FeAs$ compounds[25,29]. We cannot rule out the possibility of a pseudogap in As-based compounds, which exhibit a SDW order in its parent compound[10,11]. However, the similar phenomenology observed in LaOFeP and LaOFeAs on the 20 meV pseudogap on polycrystalline samples[29] leads us to suggest a careful re-examination once single crystals of the As-based compounds become available.

Finally, let us take a look at the Fermi surface (FS) topology (Fig. 4). Three sheets of Fermi surfaces were clearly observed: two hole pockets centered at Γ and one electron pocket centered at M. Keeping in mind the nearly degenerate $\Gamma_1$ and $M$ bands, the observed FS topology are consistent with the five sheets of Fermi surfaces predicted in band structure calculations[23]. Note that the outer hole pocket $\Gamma_2$ originates from the hybridized Fe $d_{3z^2-r^2}$ and P $p$ states, which has strong $k_z$ dispersion. The topology of this FS sheet is sensitive to the position of P atom, *i.e.,* the level of hybridization, and changes significantly upon doping. Counting the FS volume enclosed under three pockets yields 1.94, 1.03, and 0.05 electrons for $\Gamma_1$, $\Gamma_2$ and $M$ pockets, respectively. Taking into account the unresolved, nearly degenerate sheets under $\Gamma_1$ and $M$ pockets, a total electron count of $5.0 \pm 0.1$ is obtained, which is smaller than the expected value of 6. This is consistent with the need to shift $E_F$ in order to produce the best fits of the dispersion in Fig. 2. It is too early to be certain how much of this discrepancy is caused by a change in the surface doping, $k_z$ dispersion or subtle surface structure distortion, which can be significant for the $\Gamma_2$ band. As another possible discrepancy to band structure comparison, careful examination of data in Fig.3 also reveals a very weak feature around 70 meV near Γ (Fig. 3**a**), which does not seem to have corresponding band in LDA calculations. Further investigations are required to elucidate its origin. Despite these disagreements, all the expected Fermi surface pieces are observed and are in good agreement with experiments in terms of the Brillouin zone locations and signs (hole *vs.* electron). Furthermore, the measured main dispersions agree with the calculated band structures in great detail as shown in Fig. 2. These observations make a strong case that the itinerant band structure captures the essence of the electronic structure of LaOFeP.

In summary, our ARPES data from LaOFeP suggest that the electronic structure of this material can be described starting with itinerant band approach. Compared with copper oxide superconductors, it has three important contrasting features: i) it has much higher density of states near the Fermi level; ii) it has multiple bands and Fermi surface sheets; iii) it shows no apparent evidence of the pseudogap effect.

**Methods**

Single crystals of LaOFeP, with dimensions up to $0.4 \times 0.4 \times 0.04$ mm$^3$, were grown from a tin flux, using modified conditions from those first described by Zimmer and coworkers[30], and will be described elsewhere [J. Analytis, J.-H. Chu, A. S. Erickson, C. A. Cox, H. Hope, S. M. Kauzlarich, T. H. Geballe & I. R. Fisher, in preparation]. Refinements to single crystal x-ray diffraction data were consistent with previously published results, and indicated full occupancy for all sites. $T_c$, determined from resistivity ($\rho$) and susceptibility measurements, was 5.9±0.3 K. Residual resistance ratios, $\rho(300K)/\rho_0$, were up to 85, indicative of high crystal quality.

ARPES measurements were carried out at beamline 10.0.1 of the Advanced Light Source (ALS) using a SCIENTA R4000 electron analyzer. All data presented in this paper were recorded using 42.5 eV photons. The total energy resolution was set to 16 meV and the angular resolution was 0.3° (except the VB spectrum was collected using transmission mode). Single crystals were cleaved *in situ* and measured at 20 K in an ultra high vacuum chamber with a base pressure better than $2\times10^{-11}$ Torr.

Electronic structure calculations were performed within the local density approximation using the general potential linearized augmented planewave (LAPW) method. The convergence of the basis set and zone sampling was checked. Local orbitals were used to relax linearization errors and to include the semicore levels of the metal atoms. Despite the relatively low $T_c$ of LaOFeP compared with its sister compounds F-doped LaOFeAs, the calculated band structures of these two systems are very similar if one compares the calculations of LaOFeP based on experimental structure with that of LaOFeAs using optimized structure[18,23]. The calculations reported here were done using the experimental lattice parameters but relaxed internal atomic positions. There is a sensitivity of the Fermi surface to these coordinates, and as a result the present Fermi surface differs somewhat, mainly in the heavily 3D hole pocket, from that of Lebegue[23] who did not relax these coordinates. It is an interesting issue that the ARPES data on LaOFeP show better agreement with LDA band structures using optimized structure. To what extent this is due to pnictogen volatility on the cleaved surface or during the crystal growth and to what extent it indicates more subtle physics remains to be established.

**Acknowledgements** We thank C. Cox, S. M. Kauzlarich and H. Hope for single crystal x-ray diffraction measurements, and H. Yao, S. A. Kivelson, R. M. Martin, S. C. Zhang and X. L. Qi for discussions. ARPES experiments were performed at Advance Light Source, which is operated by the Office of Basic Energy Science, U.S. Department of Energy. Work at Stanford and ORNL are supported by DOE Office of Basic Energy Science, Division of Materials Science and Engineering

**Figure 1 Comparison between angle-integrated photoemission spectrum and calculated density of states. a**, Valence band spectrum of LaOFeP taken with 42.5 eV photons using transmission mode. It consists of a sharp intense peak near the Fermi level that is separated from a number of broad peaks at higher binding energy. The inset shows the valence band of LSCO for comparison. **b**, LDA density of states and projections onto the LAPW spheres. According to the LDA calculations, the near-$E_F$ states have dominant Fe $d$ character while the peaks at higher binding energy are mixtures of O $p$ states and hybridized Fe $d$ and P $p$ states. Compared with the calculated DOS, the near-$E_F$ peak has a narrower width than the calculated Fe $d$ states and is pushed closer to $E_F$, which is consistent with the band renormalization effect discussed in Fig. 2. The VB peaks at higher binding energy, on the other hand, are shifted towards higher binding energy, resulting in slightly larger total VB width.

**Figure 2 Comparison between ARPES spectra and LDA band structures along two high symmetry lines.** ARPES data from LaOFeP (image plots) were recorded using 42.5 eV photons with an energy resolution of 16 meV and an angular resolution of 0.3°. For better comparison with experimental data, the LDA band structures using the experimental lattice parameters with relaxed internal atomic positions are shifted up by ~0.11 eV and then renormalized by a factor of 2.2 (red lines). **a**, Along the $\Gamma$-X direction, two bands crossing $E_F$ can be clearly identified: one near the $\Gamma$-point ($\Gamma_1$) and one near the X-point ($\Gamma_2$). These two crossings are associated with two hole-like Fermi surface pockets centered at $\Gamma$. According to the LDA calculations, the inner pocket originates from Fe $d_{xz}$ and $d_{yz}$

bands that are degenerate at $\Gamma$, and the splitting of these two bands close to $\Gamma$ is too small to be resolved in our data. However, we do see evidence for the splitting at higher binding energy. The outer pocket is derived from the Fe $d_{3z^2-r^2}$ states that hybridize with the P $p$ and La orbitals. **b**, Along the $\Gamma$-M direction, in total three $E_F$ crossings are observed. In addition to the two crossings associated with two hole pockets, a crossing near the M-point can be observed, although the corresponding crossing in the 2$^{nd}$ zone is too weak to be seen due to the matrix element effect. This crossing is related to the electron pocket centered at M. The LDA calculations also predict two bands crossing $E_F$ around the M-point, which cannot be clearly resolved in our data.

**Figure 3 Energy distribution curves along two high symmetry lines. a**, EDC's along the $\Gamma$-X direction; and **b**, EDC's along the $\Gamma$-M direction. EDC's right at $k_F$ are highlighted in red. The leading-edge midpoints of these highlighted EDC's apparently reach $E_F$ for all bands crossing $E_F$, indicating the absence of a pseudogap within our experimental uncertainty in this system.

**Figure 4 Fermi surface maps of LaOFeP. a,** Two sets of FS mapping (unsymmetrized raw data) are overlaid: the first set covers more than one Brillouin zone while the second set taken mostly in the 2$^{nd}$ zone yields a better view of the FS pocket at the M-point, which is not well resolved in the first set due to polarization issue. The map is obtained by integrating the EDC's over an energy window of $E_F \pm 15$ meV. The red square highlights the boundary of the 1$^{st}$ zone. **b,**

Symmetrized FS map obtained by flipping and rotating the raw data shown in left panel along the high symmetry lines to reflect the symmetry of the crystal structure. Note that we use the Brillouin zone corresponding to two Fe unit cell with the M-point at $(\pi,\pi)$, which is $(\pi,0)$ in the large BZ for a simple Fe square lattice. Three sheets of Fermi surfaces, labeled as $\Gamma_1$ and $\Gamma_2$ and *M*, were clearly observed. As discussed in Fig. 2, the inner hole pocket $\Gamma_1$ observed in our data should contain two nearly degenerate sheets, same for the electron pocket around M. Therefore, our data are consistent with the five sheets of Fermi surfaces predicted in band structure calculations[23]: two hole pockets around $\Gamma$, two electron pockets around M, and one heavily 3D hole pocket centered at Z.

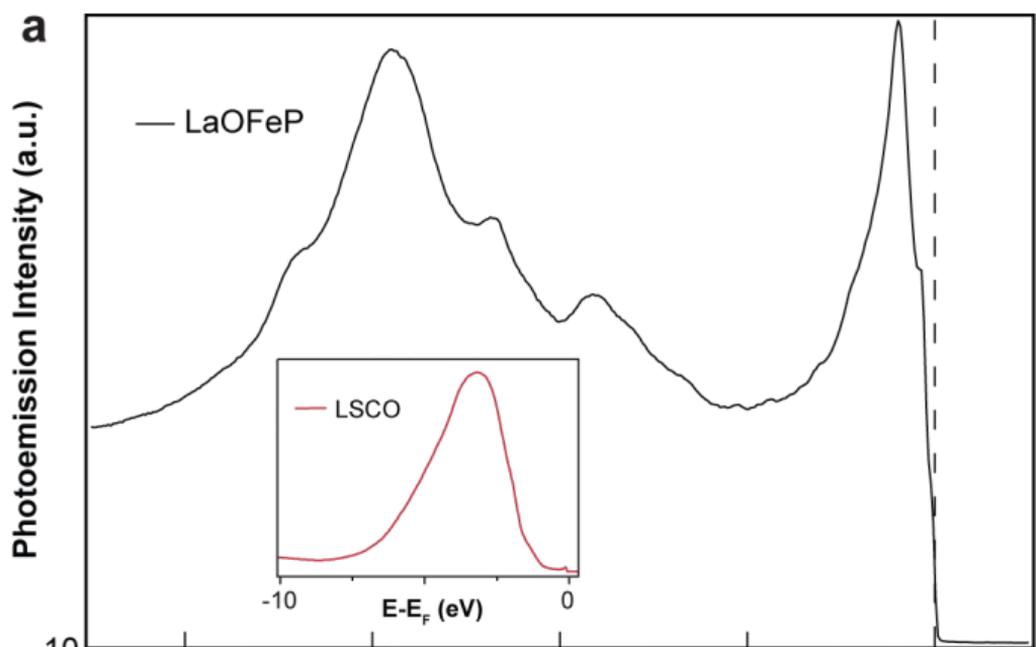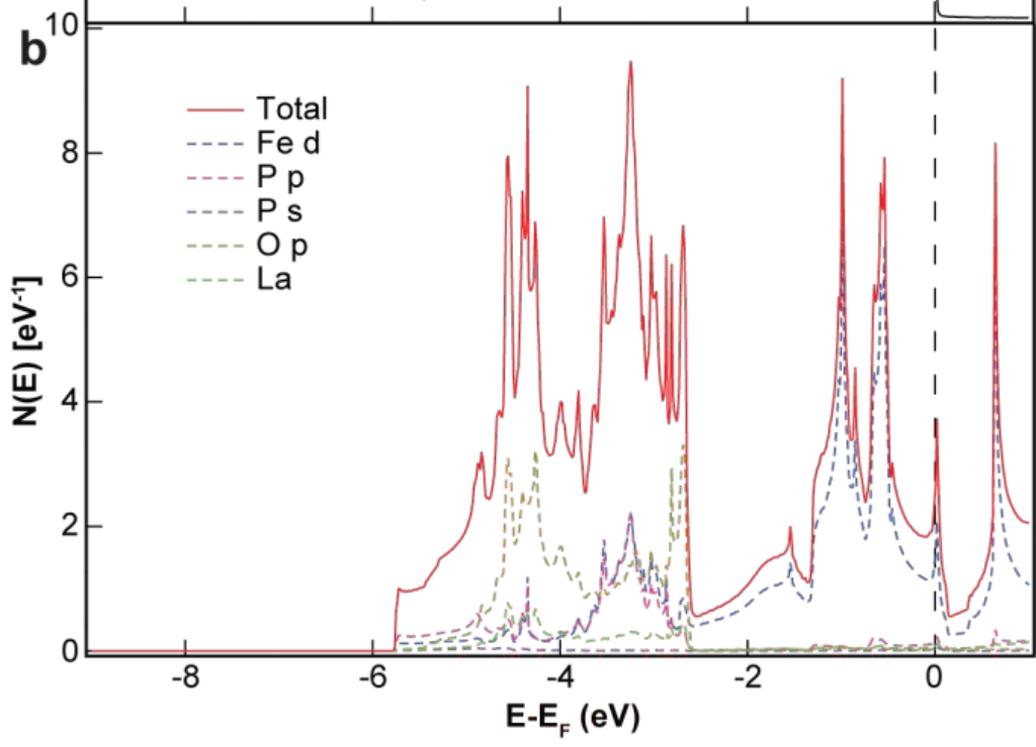

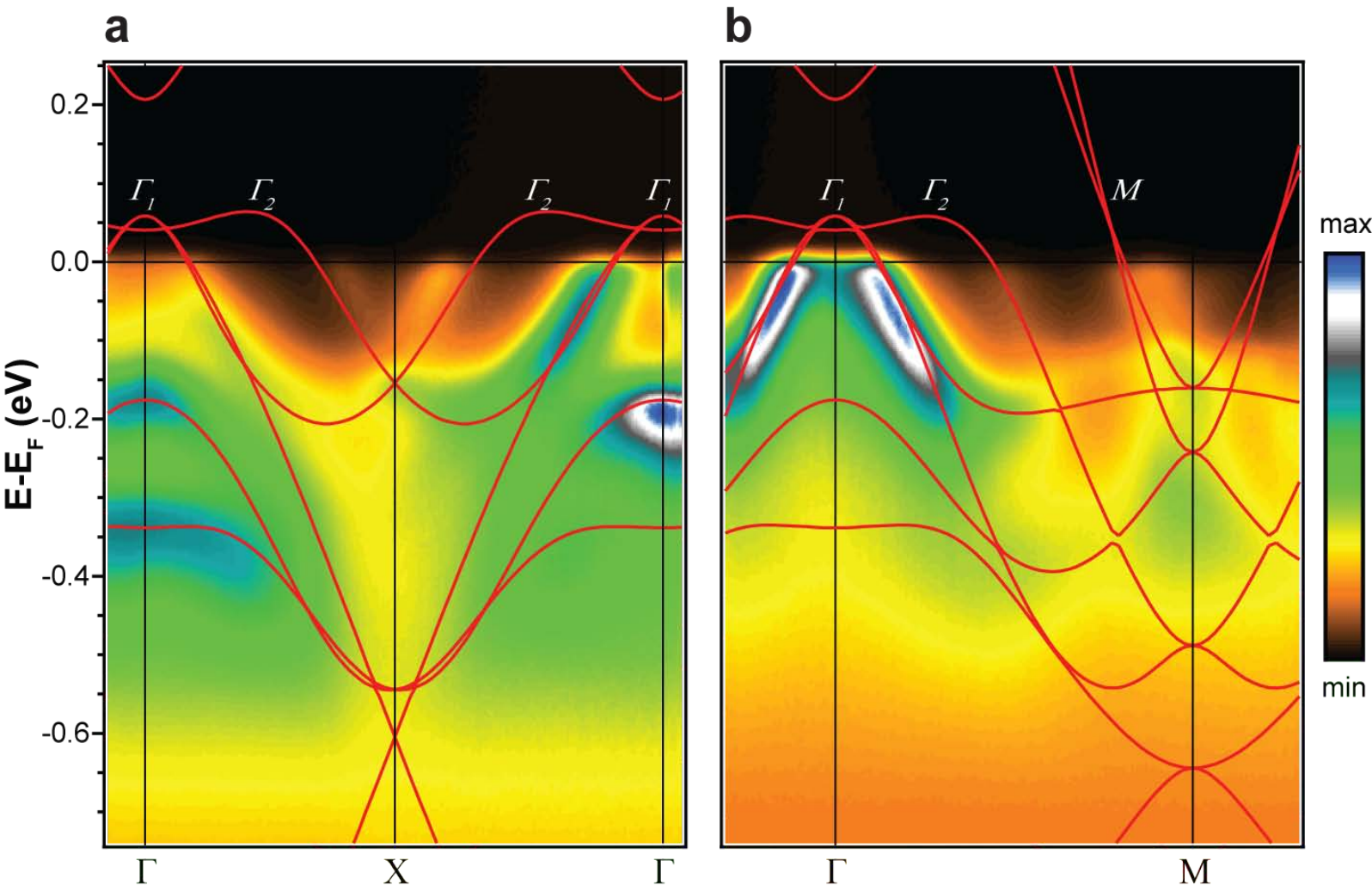

**a**

Γ

$E_1$

$E_2$

X

$E_2$

$E_1$

Γ

**b**

M

M

$M$

$E_2$

$E_1$

Γ

$E_1$

E-E$_F$ (eV)

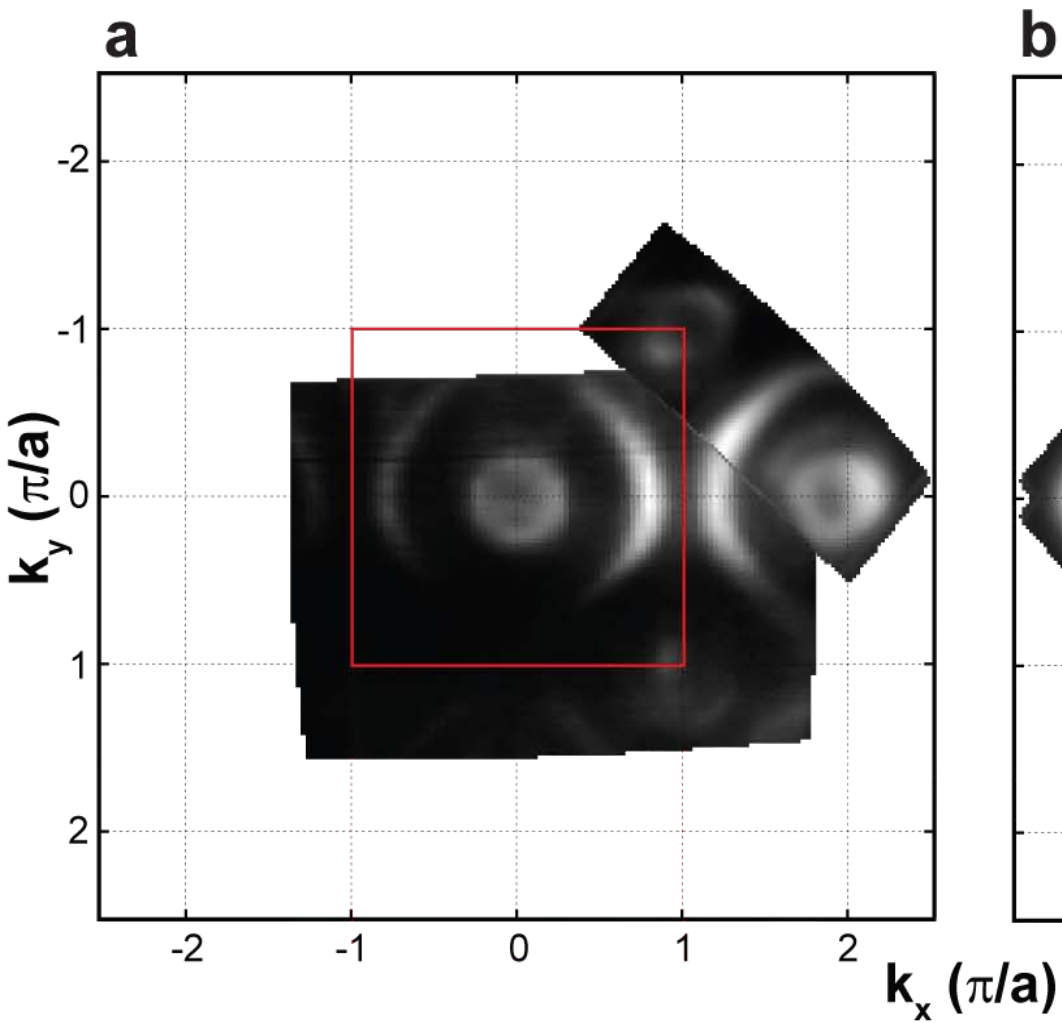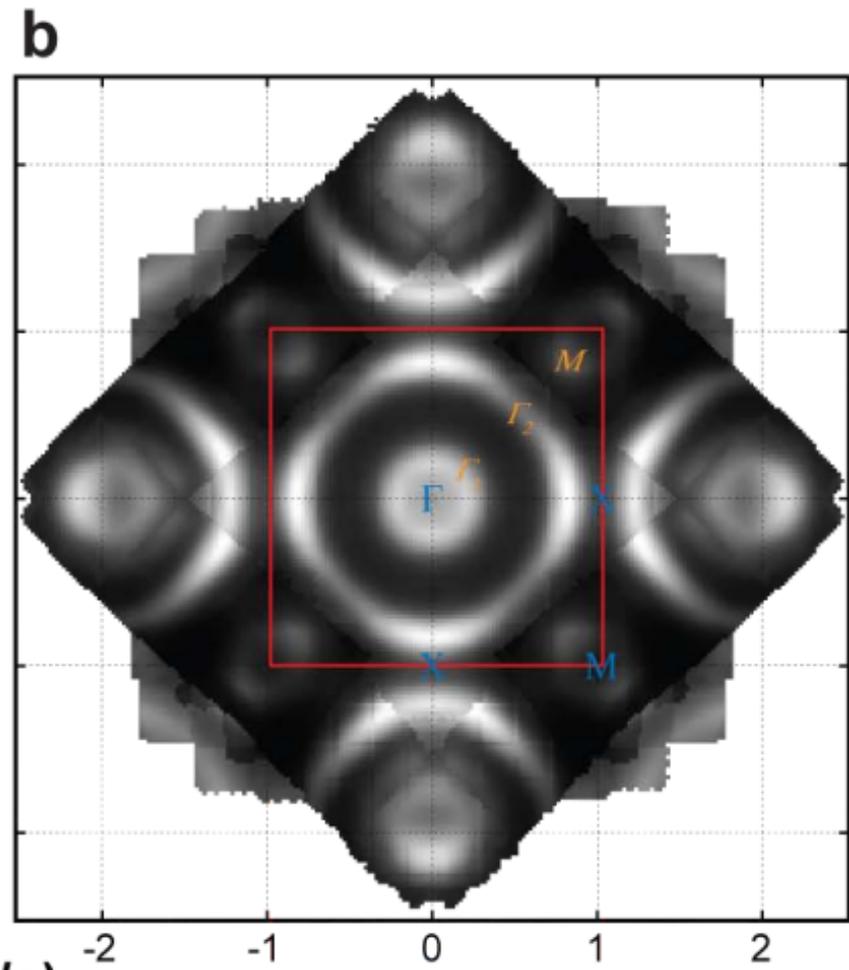